\documentclass[journal=nalefd,manuscript=article]{achemso}

\usepackage[version=3]{mhchem} 

\title{Supercurrent and multiple Andreev reflections in an InSb nanowire Josephson junction}
\author{H. A. Nilsson}
\affiliation{Division of Solid State Physics, Lund University,
P.O. Box 118, S-221 00 Lund, Sweden}
\author{P. Samuelsson}
\affiliation{Division of Mathematical Physics, Lund University,
P.O. Box 118, S-221 00 Lund, Sweden}

\author{P. Caroff}
\affiliation{I.E.M.N., UMR CNRS 8520, Avenue Poincar\'{e},  BP 60069, F-59652 Villeneuve d'Ascq, France}

\author{H. Q. Xu}
\email{Hongqi.Xu@ftf.lth.se}
\affiliation{Division of Solid State Physics, Lund University,
P.O. Box 118, S-221 00 Lund, Sweden}
\altaffiliation{Key Laboratory for the Physics and Chemistry of Nanodevices and Department of Electronics, Peking University, Beijing 100871, China}

\date{\today}

\begin{document}

\begin{abstract}
Epitaxially grown, high quality semiconductor InSb nanowires are emerging material systems for the development of high performance nanoelectronics and quantum information processing and communication devices, and for the studies of new physical phenomena in solid state systems. Here, we report on measurements of a superconductor-normal conductor-superconductor junction device fabricated from an InSb nanowire with aluminum based superconducting contacts. The measurements show a proximity induced supercurrent flowing through the InSb nanowire segment, with a critical current tunable by a gate, in the current bias configuration and multiple Andreev reflection characteristics in the voltage bias configuration. The temperature dependence and the magnetic field dependence of the critical current and the multiple Andreev reflection characteristics of the junction are also studied. Furthermore, we  extract the excess current from the measurements and study its temperature and magnetic field dependences. The successful observation of the superconductivity in the InSb nanowire based Josephson junction device indicates that InSb nanowires provide an excellent material system for creating and observing novel physical phenomena such as Majorana fermions in solid state systems. 

Keywords: InSb nanowires, Josephson junction, supercurrent, multiple Andreev reflection
\end{abstract}

\maketitle

Epitaxially grown semiconductor nanowires are emerging nanomaterials with potential applications in nanoelectronics and optoelectronics.\cite{Xia,Huang,Duan,Gudiksen,Duan2,Johnson,Hua,Tian,Kelzenberg,Dong,Cao,Boxberg,Xiang,Yan,Nguyen,Ng,Goldberger,Schmidt,Bryllert,Nilsson} Novel wrap-gate field-effect transistors,\cite{Nguyen,Ng,Goldberger,Schmidt,Bryllert} light emitting devices,\cite{Duan,Gudiksen,Duan2,Johnson,Hua} and photovoltaic devices\cite{Tian,Kelzenberg,Dong,Cao,Boxberg} have been proposed and developed with these high crystalline-quality nanowires. These nanowires are also highly desirable nanomaterials for realizing novel physical systems in the parameter space suitable for the investigations of physical phenomena arising from the strong quantum confinement, Coulomb interaction, and spin correlation.\cite{Roddaro,Nilsson2,Nilsson3,Nadj,Kristinsdottir} Here, the intrinsic one-dimensional geometrical structure of the nanowires has greatly simplified the device fabrication process. In addition, a variety of metal contact combinations can be employed, opening up great possibilities in the realization of hybrid devices\cite{Doh,Dam,Xiang2,Jespersen,Hofstetter,Franceschi} for new physics studies. Among them, epitaxially grown, high crystalline-quality InSb nanowires\cite{Nilsson2,Caroff,Lugani} are one of most interesting material systems, due to the fact that bulk InSb material~\cite{Vurgaftman,Isaacson}  has a very high electron mobility $\mu_e \sim 77000$ cm$^2$/Vs, a very small electron effective mass $m_e^* \sim 0.015$ m$_e$, and a very large electron magnetic moment $|g^*| \sim 51$. Recently, quantum dot devices manufactured from InSb nanowires have shown to exhibit giant, strongly level-dependent electron g-factors and strong spin-orbit interactions.~\cite{Nilsson2} As a result, InSb nanowires have been suggested as one of most promising material systems for creating novel particles--Majorana fermions--in solid state systems.\cite{Kitaev,Lutchyn,Oreg,Akhmerov,Flensberg2011,Lutchyn2,Stanescu} However, one of the prerequisites for creating Majorana fermions in an InSb nanowire based device is to introduce superconductivity in the nanowire device by the proximity effect. Thus, an experimental verification of the proximity induced superconductivity in the InSb nanowires is an important step towards realizing Majorana fermions in solid state structures. 

Here, we report a first experimental study of a superconductor-normal conductor-superconductor (S-N-S) junction device fabricated from the InSb nanowire segment of an InSb/InAs nanowire heterostructure with aluminum based superconducting contacts. The measurements show a proximity induced supercurrent flowing through the InSb nanowire junction, with a critical current tunable by a gate, and multiple Andreev reflection (MAR) characteristics. The temperature dependence and the magnetic field dependence of the critical current and the MAR characteristics of the InSb nanowire based S-N-S junction device are also studied in details. Furthermore, we extract the excess current from the measurements and study its temperature and magnetic field dependences. The successful observation of the proximity induced superconductivity in the S-N-S junction device indicates that InSb nanowires provide an excellent solid state system for creating and observing Majorana fermions.

\begin{figure}[t]
  \begin{center}
    \includegraphics[width=8.5cm]{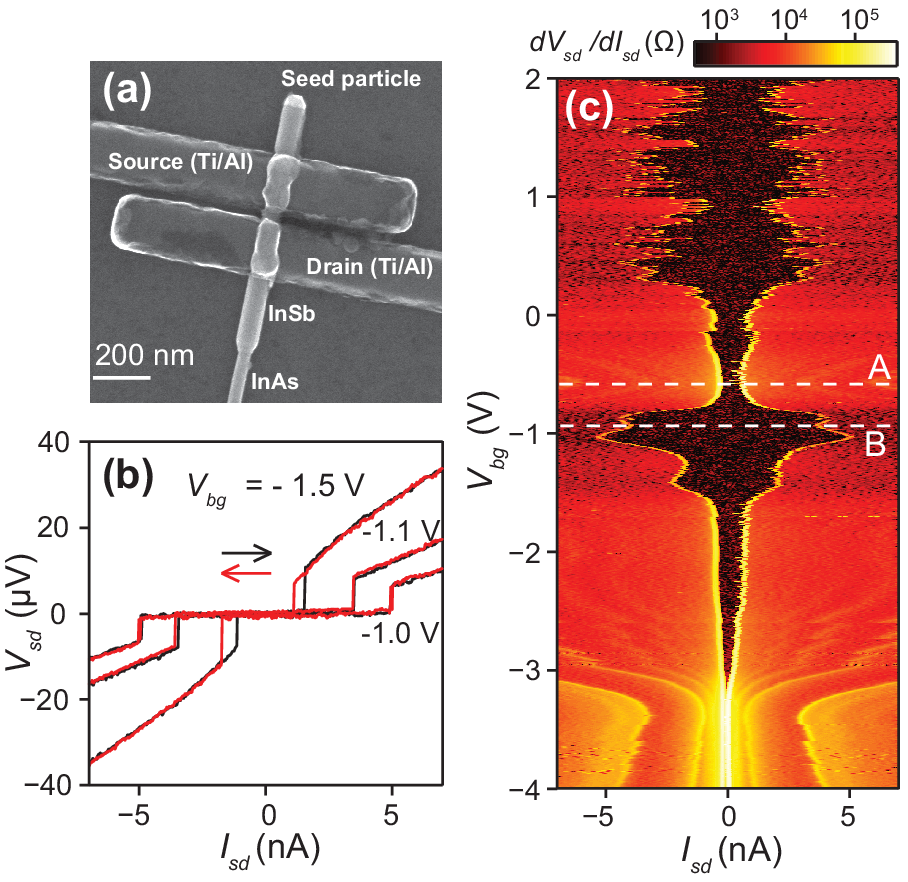}
  \end{center}
  \caption{(a) SEM image of an InSb nanowire based S-N-S junction device. The device is made from the InSb segment of an InAs/InSb heterostructure nanowire with Ti/Al metal contacts, on a SiO$_2$ capped, highly doped Si substrate, using electron beam lithography. (b) Source-drain voltage $V_{sd}$ measured for the device as a function of the source-drain current $I_{sd}$ at three different gate voltages $V_g = -1.5$, $-1.1$, and $-1.0$ V, and at base temperature $T = 25$ mK. The black curves are recorded in the upward current sweeping direction and the red curves in the downward current sweeping direction. The hysteresis is seen in the $V_{sd}$-$I_{sd}$ characteristics of the device measured at $V_{bg}=-1.5$ V. A supercurrent is seen to flow through the InSb nanowire segment, due to superconductivity induced by the proximity effect, in the regions with negligible source-drain voltages. A critical current $I_c$ can be identified as an applied current at which the measured source-drain voltage jumps to a finite value. (c) Differential resistance $dV_{sd}/dI_{sd}$, on a color scale, as a function of the source-drain current $I_{sd}$ and the silicon back gate voltage $V_{bg}$, measured at base temperature $T = 25$ mK. The central dark area corresponds to a region with negligible device resistance (superconductivity region). The critical current $I_c$, which can be identified from the edge of the dark area, varies strongly with $V_{bg}$.}
  \label{figure1}
\end{figure}

The InSb nanowire based S-N-S junction devices investigated here are fabricated from InSb segments of InSb/InAs heterostructure nanowires. The heterostructure nanowires are grown on InAs(111)B substrates by metal-organic vapor phase epitaxy in a two-stage approach using aerosol gold particles with a diameter of 40 nm as initial seeds.\cite{Nilsson2,Caroff} In the first stage of growth, the InAs nanowire segments are grown from the gold particles using precursor materials of trimethylindium (TMIn) and arsine. In the second stage of growth, arsine is switched to trimethylantimony (TMSb) and the grown InAs segments serve as seeds to favor nucleation of InSb. Moreover, the grown InAs segments have ensured physical decoupling of the top InSb segment growth and substrate surface growth, thus favoring stable and reproducible growth of InSb nanowires. At the heterostructure interfaces, the diameter of the nanowires increases substantially after the precursor switch due to an increase in the indium content in the metal particles. Thus, the grown InSb nanowire segments have a larger diameter than the InAs nanowire segments, as clearly visible in \ref{figure1}a. For further details about the growth and structural properties of the InAs/InSb heterostructure nanowires  employed in this work, we refer to Refs.~22 and 32. 

The grown InAs/InSb heterostructure nanowires are transferred to a SiO$_2$ capped, degenerately doped, n-type Si substrate with predefined Ti/Au bonding pads and metal markers. Using an optical microscope, the wire positions relative to the metal markers are recorded. The substrate is then spin-coated with poly(methyl methacrylate) (PMMA) resist. Using electron beam lithography, two 200-nm-wide contact areas with a varying spacing between them are defined on the InSb segment of each selected heterostructure nanowire. To obtain a clean metal-semiconductor interface, the exposed semiconductor contact areas are briefly etched in a (NH$_4$)$_2$S$_x$ solution, followed by a rinse in H$_2$O before metal deposition. Finally, metal contact electrodes are defined by vapor deposition of 5-nm-thick Ti and 85-nm-thick Al metal, and a process of lift-off in hot acetone. A scanning electron microscope (SEM) image of a fabricated device is shown in \ref{figure1}a. The room-temperature resistance of fabricated devices is measured using a probe station and the devices with room-temperature resistance in the order of 10 k$\Omega$ are selected for detailed studies at low temperatures. In this work, we report the results of low-temperature electrical measurements of a selected device as shown in \ref{figure1}a with a spacing of 30 nm between the two superconductor contacts--a device among the shortest nanowire-based S-N-S junctions. All electrical measurements presented below are performed in a $^3$He/$^4$He dilution refrigerator.

\ref{figure1}b shows the V-I characteristics of the device at a cryostat base temperature of 25 mK. Here the source-drain voltage $V_{sd}$ is measured as a function of the applied source-drain current $I_{sd}$ at back gate voltages $V_{bg} = -1.5$, $-1.1$, and $-1.0$ V.  In this figure, the measured voltages for both current sweep directions are plotted, where the black curves are for the upward current sweeps and the red curves are for the downward current sweeps.  The measured V-I characteristics exhibit a clear superconductive branch (i.e., a zero resistance region) and a dissipative quasiparticle branch. The appearance of the dissipationless supercurrent in the device is a clear manifestation of the proximity effect, which can be viewed as a consequence of the diffusion of Cooper pairs through the entire nanowire section between the two superconducting contacts. The switching between the superconductive and the dissipative branch depends on the back gate voltage and the sweeping direction of the source-drain current. At a given back gate voltage, the switching from superconductive to dissipative conduction occurs when the current approaches a critical value $I_c$, leading to an abrupt appearance of a finite voltage. The reversed switching from resistive to superconductive state may not occur at the same critical value of the current, but a lower current level $I_r$, as seen in the measured V-I curve at the back gate voltage $V_{bg}=-1.5$ V. This hysteretic behavior, which has also been seen in Josephson junctions made from InAs nanowires\cite{Doh} and from Si/Ge nanowires,\cite{Xiang2} could be the result of phase instability typically found in a capacitively and resistively shunted Josephson junction or simply due to a heating effect.\cite{Tinkham,Tinkham2} 

\ref{figure1}c displays the differential resistance $dV_{sd}/dI_{sd}$, on a color scale, as a function of source-drain current $I_{sd}$ and back gate voltage $V_{bg}$ at the base temperature of $T = 25$ mK. Here, only the data recorded in the upward current sweeps are used in deriving the differential resistance $dV_{sd}/dI_{sd}$ and thus the $dV_{sd}/dI_{sd}$ plot is slightly asymmetric with respect to the zero current at some back gate voltages, due to the hysteretic effect discussed above. The dark central area in \ref{figure1}c corresponds to a superconductive region with negligibly low device resistance and the critical current $I_c$ can be identified at the right edge of the dark area. It is clearly seen that the critical current $I_c$  is a function of $V_{bg}$. This is a unique feature for a Josephson junction with the junction link made from a semiconductor, since the critical current $I_c$ is related to the resistance $R_n$ of the link in the normal state\cite{Tinkham,Beenakker} as $I_c R_n \sim  \Delta /e$ at temperature well below $T_c$ and the resistance $R_n$ can be tuned by a gate in a semiconductor based junction device. Also, there exist two distinct regions of back gate voltages in which the critical current $I_c$ is large: the region around $V_{bg}=-1$ V and the region with $V_{bg}> 0$ V. In the region around $V_{bg}=-1$ V, the critical current $I_c$ shows a smooth variation and in the other region the critical current $I_c$ shows rapid fluctuations as a function of $V_{bg}$. 

\begin{figure}[t]
  \begin{center}
    \includegraphics[width=8.5cm]{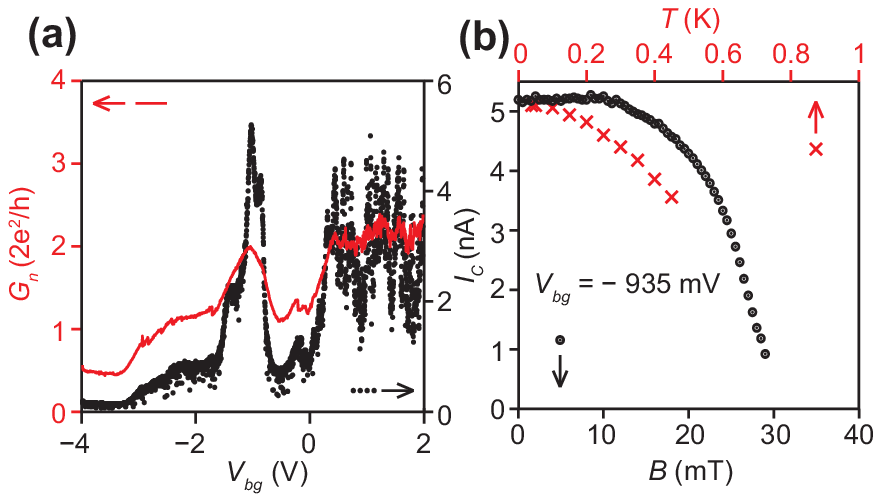}
  \end{center}
  \caption{(a) Normal state conductance $G_n$ (red solid curve) and critical current $I_c$ (black dots) measured against back gate voltage $V_{bg}$ at base temperature $T = 25$ mK. The normal state conductance in the linear response regime is measured at magnetic field $B = 100$ mT larger than the critical field $B_c$ of the superconductor Al. The measurements show that the $I_cR_n$ product is not a constant in the device, but varies between 4 $\mu$V and 34 $\mu$V. (b) Critical current $I_c$ measured as a function of magnetic field $B$ (black dots) and as a function of temperature $T$ (red crosses) at back gate voltage $V_{bg} = -935$ mV (corresponding to cut B in \ref{figure1}a). With increasing magnetic field $B$, $I_c$ remains constant up to $B\sim 10$ mT and then decreases smoothly with a critical magnetic field of $B_c \sim 30$ mT. With increasing temperature $T$, $I_c$ decreases instantly. But at temperatures above $T = 0.45$ K, $I_c$ can no longer be identified from the measurements.}
  \label{figure2}
\end{figure}

To understand the appearance of the distinct characteristics of the critical current $I_c$ in the two back gate voltage regions, we measure the normal state conductance $G_n$ of the InSb nanowire junction device and plot the results (red solid curve) along with the measured critical current $I_c$ (black dots) in \ref{figure2}a.  The measurements of the normal state conductance $G_n$ are made at $T = 25$ mK by applying a magnetic field $B = 100$ mT, sufficiently larger than the critical magnetic field of the Al electrodes, to the device in the linear response regime. It is seen in \ref{figure2}a that the normal state conductance $G_n$ of the InSb nanowire junction shows a broad resonant peak on a conductance background of about $2e^2/h$ with the peak value close to $4e^2/h$ at $V_{bg} \sim -1$ V and rapid fluctuations around $4e^2/h$ at $V_{bg}>0$ V. 
The resonant peak can be attributed to transport through a broadened energy level in the InSb nanowire junction, while the rapid fluctuations resemble mesoscopic conductance fluctuations of a coherent, quasi-ballistic or diffusive conductor, indicating that the transport in the InSb nanowire section is quasi-ballistic or diffusive at $V_{bg}>0$ V. In addition, the overall measured normal state conductance values are in the range of $2e^2/h$ to $4e^2/h$ at the back gate voltages of $-3$ V to 2 V, implying that the normal state transport in the InSb nanowire section is in the few-channel regime in this back gate voltage range.
\ref{figure2}a also clearly shows that the Josephson critical current $I_c$ of the device is correlated closely to the normal state conductance $G_n$ of the InSb nanowire junction. Furthermore, it is seen that the relative amplitudes of the critical current fluctuations $\delta I_c/I_c$ are much larger than the relative amplitudes of the normal state conductance fluctuations $\delta G_n/G_n$. However, it should be noted that the measured $I_cR_n$ product is not a constant in our device, but shows a variation between 4 $\mu$V and 34 $\mu$V. These values overall are significantly lower than the value of $I_cR_n \sim \Delta/e = 150$ $\mu$V expected for an ideal S-N-S junction embedding a short and diffusive normal conductor. Such reduced experimental $I_cR_n$ values compared to theory were also found in InAs nanowire based S-N-S junctions\cite{Doh} and can typically be attributed to premature switching due to thermal activation in a capacitively and resistively shunted junction and to finite transparency of the superconductor contacts to the InSb nanowire.\cite{Tinkham}

\ref{figure2}b shows the influences of magnetic field $B$ and temperature $T$ on the critical current $I_c$ of the InSb nanowire based S-N-S junction device. Here, the critical current $I_c$ measured at back-gate voltage $V_{bg} = -935$ mV (corresponding to cut B in \ref{figure1}a) is plotted as a function of magnetic field $B$ (black dots) and as a function of temperature $T$ (red crosses). With increasing magnetic field, the critical current $I_c$ remains constant up to $B \sim 10$ mT, followed by a smooth decrease with a critical magnetic field of $B_c \sim 30$ mT. With increasing temperature, the critical current $I_c$ decreases instantly towards smaller values. However, at temperatures above $T \sim 0.45$ K, the critical current $I_c$ can no longer be identified. Overall the critical current $I_c$ is related to the change in the superconducting energy gap $\Delta$ with increasing B or T as described in Ref.~43.

\begin{figure}[t]
  \begin{center}
    \includegraphics[width=8.5cm]{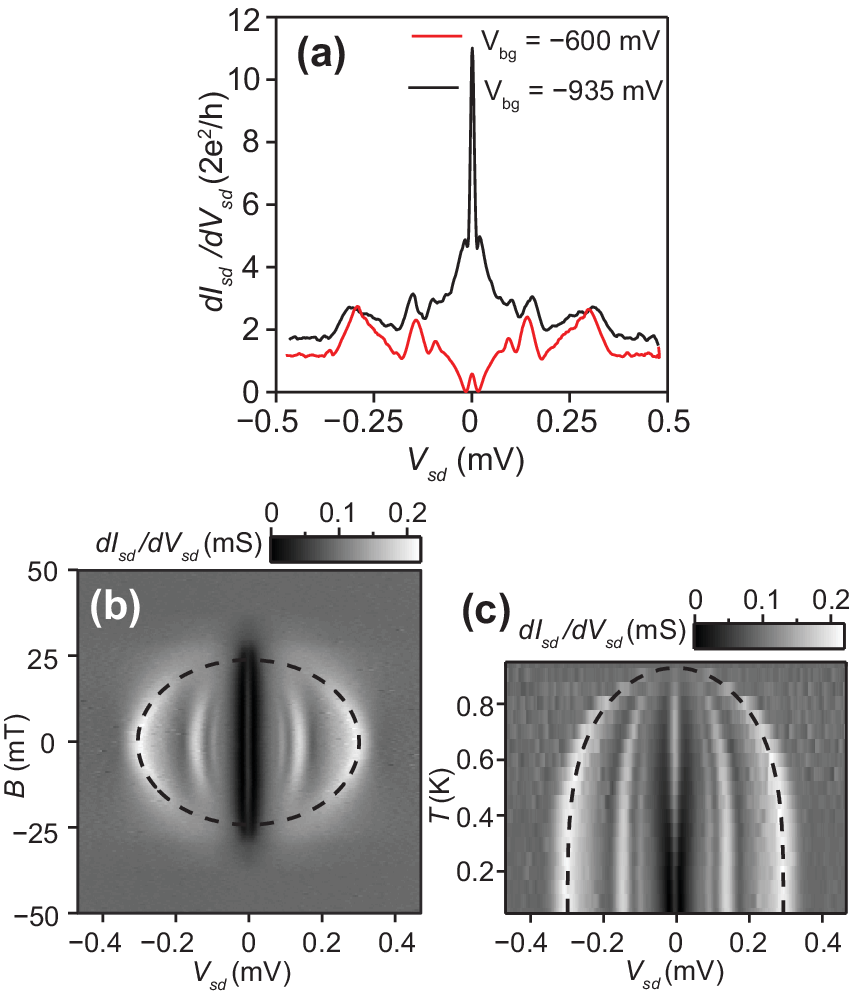}
  \end{center}
  \caption{(a) Differential conductance $dI_{sd}/dV_{sd}$ measured as a function of source-drain voltage $V_{sd}$ at back gate voltage $V_{bg} = -935$ mV (black solid curve, corresponding to cut B in \ref{figure1}c) and $V_{bg} = -600$ mV (red solid curve, corresponding to cut A in \ref{figure1}c) at base temperature $T = 25$ mK. Both curves are seen to display a series of peaks, due to MARs, symmetric around $V_{sd} = 0$ V. The two curves also show different behaviors in the magnitude of $dI_{sd}/dV_{sd}$ as $|V_{sd}|$ is decreased to zero. The black solid curve shows that $dI_{sd}/dV_{sd}$ increases on average, followed by a strong supercurrent peak at $V_{sd} = 0$ V, while the red solid curve shows that $dI_{sd}/dV_{sd}$ decreases on average, followed by a strongly suppressed supercurrent peak at $V_{sd} = 0$ V. (b) Differential conductance $dI_{sd}/dV_{sd}$, on a gray scale, measured as a function of source-drain bias voltage $V_{sd}$ and magnetic field $B$ at base temperature $T = 25$ mK and gate voltage $V_{bg} = -600$ mV. Here, the peaks in $dI_{sd}/dV_{sd}$ are seen as bright white lines. The black dashed line indicates the theoretically predicted magnetic field dependence of the superconducting energy gap of the Al based electrodes. (c) Differential conductance $dI_{sd}/dV_{sd}$, on a gray scale, measured as a function of source-drain bias voltage $V_{sd}$ and base temperature $T$ at zero magnetic field and back gate voltage $V_g = -600$ mV. The black dashed curve displays the theoretically predicted temperature dependence of the superconducting energy gap of the Al based electrodes.}
  \label{figure3}
\end{figure}

We now turn to investigate the characteristics of the InSb nanowire based S-N-S junction device in the voltage bias configuration. \ref{figure3}a displays the differential conductance $dI_{sd}/dV_{sd}$ of the junction device as a function of source-drain voltage $V_{sd}$ at the base temperature of $T = 25$ mK and the zero magnetic field. The red solid curve in the figure represents the measurements at back gate voltage $V_{bg} = -600$ mV (corresponding to cut A in \ref{figure1}a) at which the device shows a small $I_c$, while the black solid curve in the figure shows the measurements at back gate voltage $V_{bg} = -935$ mV (corresponding to cut B in \ref{figure1}a) at which the device exhibits a large $I_c$. It is seen that both curves display a series of peaks, symmetric around $V_{sd} = 0$ V. These peaks can be attributed to MAR processes. It has been shown that MAR processes in an S-N-S junction give rise to a series of structures in the differential conductance $dI_{sd}/dV_{sd}$ at source-drain voltages $V_{sd} = 2\Delta/en$, where $n = 1, 2, 3, ...$ corresponds to the number of carrier transfers across the junction.\cite{Klapwijk,Octavio,Flensberg,Kleinsasser}   In \ref{figure3}a, the first three peaks are clearly visible at the voltages of $V_{sd} \sim 2\Delta/en$ with $n = 1$ to 3. From a comparison with the theory for coherent MARs in short S-N-S junctions,\cite{Bratus,Averin,Cuevas} we can deduce a value of $\Delta = 150$ $\mu$V. Note that $\Delta$ determined in this way is slightly different at different back gate voltages. Also, in both curves, a sharp peak at $V_{sd} = 0$ V is observed, due to the superconductivity induced by proximity effect in the InSb nanowire section. However, comparing the two curves in \ref{figure3}a, we observe that the central peak is much more pronounced at $V_{bg} = -935$ mV (black solid curve), at which the critical current $I_c$ is large. We also observe that the overall trend of $dI_{sd}/dV_{sd}$ with decreasing $|V_{sd}|$ differs distinctly in the two curves. At $V_{bg} = -935$ mV (black solid curve) $dI_{sd}/dV_{sd}$ increases as $|V_{sd}|$ decreases to zero, whereas at $V_{bg} = -600$ mV (red solid curve) $dI_{sd}/dV_{sd}$ decreases. This behavior is consistent with the our early discussion that transport through a broadened energy level in the InSb nanowire junction plays an important role in this back gate voltage range. The suppression of higher order ($n\geq 4$) MAR features in the differential conductance can be attributed to inelastic processes and  loss of the carrier coherence in the multiple carrier transfers across the junction.

We now proceed to investigate the magnetic field and temperature dependences of the MAR features in the differential conductance of the S-N-S junction device. \ref{figure3}b displays the differential conductance $dI_{sd}/dV_{sd}$, on a gray scale, as a function of source-drain bias voltage $V_{sd}$ and magnetic field $B$ at the gate voltage $V_{bg} = -600$ mV (corresponding to cut A in \ref{figure1}a). Here, we observe the peaks of high differential conductance as bright lines. As the magnetic field is increased from zero, the peaks are smoothly shifted toward lower $|V_{sd}|$, as the superconducting energy gap of the Al based electrodes is decreased. The black dashed line indicates the theoretically predicted magnetic field dependence of the superconducting energy gap $2\Delta(B)/e$, where $\Delta(B)=\Delta\sqrt{1-(B/B_c)^2}$ and $B_c$ is the critical magnetic field of the superconductor. We note that here the measured peak positions of the differential conductance deviate slightly from the theoretical values of $\Delta(B)$ at large $B$. \ref{figure3}c shows the differential conductance $dI_{sd}/dV_{sd}$ as a function of source-drain bias voltage $V_{sd}$ and temperature $T$ at zero magnetic field and again the back gate voltage of $V_{bg} = -600$ V. As the temperature is increased, the MAR peaks in the differential conductance again move toward lower $|V_{sd}|$, corresponding to a decrease of the superconducting energy gap of the Al based electrodes. The black dashed line represents the theoretically predicted temperature dependent superconducting energy gap $2\Delta (T)/e$, where $\Delta(T)=\Delta\sqrt{\cos[(\pi/2)(T/T_c)^2]}$ and $T_c$ is the critical temperature of the superconductor. From the temperature-dependent measurements, we deduce a zero-temperature superconducting energy gap $\Delta\sim 148$ $\mu$V and critical temperature $T_c \sim 0.92$ K for the Al based electrodes. 

\begin{figure}[t]
  \begin{center}
    \includegraphics[width=8.5cm]{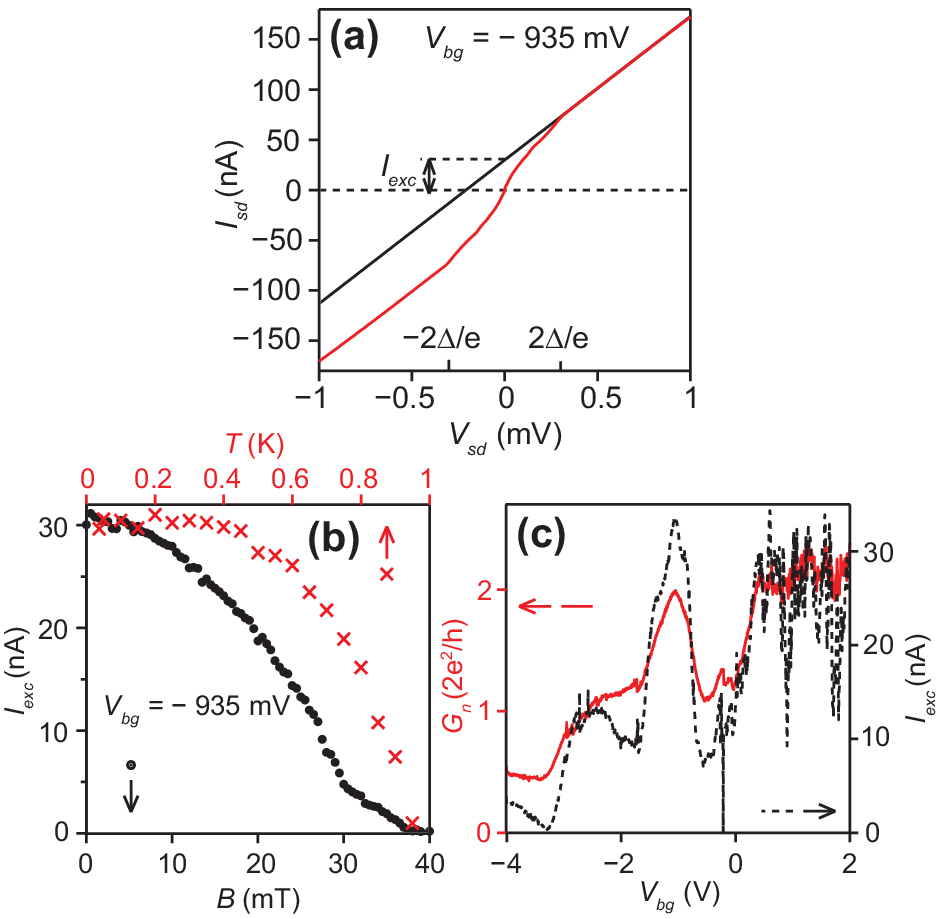}
  \end{center}
  \caption{(a) Source-drain current $I_{sd}$ (red solid curve) as a function of source-drain bias $V_{sd}$ measured at base temperature $T = 25$ mK and back gate voltage $V_{bg} = -935$ mV (corresponding to cut B in \ref{figure1}a). The current $I_{sd}$ at large $V_{sd} > 2\Delta/e$ is linear in $V_{sd}$. However, it does not extrapolate back to zero at $V_{sd} = 0$ V, but to a finite excess current $I_{exc}$. (b) Excess current $I_{exc}$ as a function of temperature $T$ (red crosses) and as a function of magnetic field $B$ (black dots) extracted for the device at back gate voltage $V_{bg} = -935$ mV (corresponding to cut B in \ref{figure1}a). (c) Normal state conductance $G_n$ (red solid curve) and excess current $I_{exc}$ (black dots) plotted against back gate voltage $V_{bg}$ for the device at base temperature $T = 25$ mK. The normal state conductance is measured in the linear response regime at magnetic field $B = 100$ mT $> B_c$ (the same as in \ref{figure2}a). The $I_{exc}R_n$ product extracted from the measurements varies between 60 $\mu$V and 220 $\mu$V.}
  \label{figure4}
\end{figure}

\ref{figure4}a displays the source-drain current $I_{sd}$ (red solid curve) measured as a function of source-drain voltage $V_{sd}$ at back gate voltage $V_{bg} = -935$ mV, zero magnetic field, and temperature $T = 25$ mK. The measurements show clearly that the $I_{sd}-V_{sd}$ curve at the source-drain voltages of $V_{sd} > 2\Delta_0/e$ does not extrapolate back to $I_{sd}=0$ at $V_{sd} = 0$, but to a finite excess current $I_{exc}$ (see the black solid line in the figure). The excess current, formally the difference between $I_{sd}$ and the normal state current at $V_{sd}>>\Delta/e$, arises due to an interplay of currents carried by particles undergoing zero and one Andreev reflections when traversing the junction at large bias voltages.\cite{Bratus,Averin,Cuevas} The excess current is of particular interest, since it contains information on the junction properties and is, in contrast to $I_c$, practically insensitive to decoherence for transparent S-N-S junctions. \ref{figure4}b shows the behavior of $I_{exc}$ as a function of temperature $T$ (red crosses) and as a function of magnetic field $B$ (black dots). With increasing $T$, the excess current $I_{exc}$ first decreases slowly up to $T \sim 0.4$ K and then decreases quickly until it reaches zero at a temperature of $T\sim 0.95$ K. With increasing $B$, the excess current $I_{exc}$ decreases smoothly towards zero. Theory\cite{Klapwijk,Cuevas} predicts that $I_{exc} \propto \Delta$ at a given $V_{bg}$ and, therefore, the observed decrease of $I_{exc}$ with increasing $B$ or $T$ is well explained by the corresponding decrease in $\Delta$. However, we note that at $B \sim 30$ mT, a kink in $I_{exc}$ is observed with a following tail of finite $I_{exc}$ values up to $B \sim 39$ mT. A weak kink is actually also visible in the measured critical current $I_c$ as a function of $B$ shown in \ref{figure2}b. The origin of this kink is not clear.
Finally, in \ref{figure4}c, we compare the behaviors of the excess current $I_{exc}$ (black dashed curve) and the normal state conductance $G_n$ (red solid curve) with change in back gate voltage $V_{bg}$. As in the case for the $I_cR_n$ product, the $I_{exc}R_n$ product is also not a constant. Here, we find a variation of $I_{exc}$ from 0.4 nA to 34 nA, yielding an $I_{exc}R_n$ product between 60 $\mu$V and 220 $\mu$V with a peak value at the back gate voltage close to $V_{bg}=-1$ V. Our measured $I_{exc}R_n$ product values are clearly much smaller than the value of $8\Delta/3e\sim 400$ $\mu$V predicted based on a short ballistic weak link model\cite{Klapwijk}, but are more comparable to the value of $(\pi^2/4 - 1) \Delta /e\sim 220$ $\mu$V predicted based on a short diffusive weak link model.\cite{Artemenko} 

In summary, we have reported a study of an InSb nanowire based S-N-S junction device and observed a proximity induced supercurrent and multiple Andreev reflections in the device. The critical current $I_c$ tunable by a voltage applied to the back gate with a maximum value of $I_c=5$ nA is observed. However, the $I_cR_n$ product is not a constant, but varies between $4$ $\mu$V and $34$ $\mu$V, substantially lower than the expected value of the device, $I_cR_n \sim \Delta/e = 150$ $\mu$V, in the coherent, diffusive transport regime. We have also observed clear MAR structures, up to third order ($n = 3$), in the measured differential conductance of the junction device and deduced a superconducting energy gap of $\Delta \sim 150$ $\mu$eV from the measurements. We have found that the behavior of $dI_{sd}/dV_{sd}$ in the low source-drain bias voltage region depends on the normal state resistance $R_n$. At a small $R_n$ the differential conductance $dI_{sd}/dV_{sd}$ is overall increased as $|V_{sd}|$ is decreased toward zero, whereas at a large $R_n$ it is decreased. We have measured the $V_{sd}$ positions of the MAR structures as a function of the magnetic field and as a function of the temperature, and have found that both are in a good agreement with the theoretical predictions. Finally, we have examined the excess current $I_{exc}$ and found that $I_{exc}$ is in the range between 0.4 nA and 34 nA, yielding an $I_{exc}R_n$ product in the range between 60 $\mu$V to 220 $\mu$V, in the back gate voltage range of $V_{bg}=-4$ V to $2$ V. The observed superconductivity in the InSb nanowire based S-N-S junction, together with previously demonstrated large spin-orbit interaction strength and large Lande g-factor in the InSb nanowire quantum structures,\cite{Nilsson2} opens up new opportunities in the design and realization of novel quantum devices and in the studies of new physics such as Majorana fermions in solid state systems. 

This work was supported by the Swedish Research Council (VR),
the Swedish Foundation for Strategic Research (SSF) through the Nanometer Structure Consortium at Lund University (nmC@LU), the National Basic Research Program of the Ministry of Science and Technology of China (Nos.
2012CB932703 and 2012CB932700).

\end{document}